\documentstyle[prl,aps,epsf,multicol,amssymb]{revtex}

\def\beginwide{
        \end{multicols} \vspace*{-0.5cm} \noindent
        \rule{3.5in}{.1mm}\rule{.1mm}{5mm} \widetext \medskip }
\def\beginwidetop{
        \end{multicols} \vspace*{-0.5cm} \noindent
        \widetext \medskip }
\def\endwide{
        \hspace*{3.5in}~\rule[-5mm]{.1mm}{5mm}\rule{3.5in}{.1mm}
        \begin{multicols}{2} \vspace*{-1.0cm} \noindent }
\def\endwidebottom{
        \begin{multicols}{2} \vspace*{-1.0cm} \noindent }

\begin{document}
\advance\textheight by 0.5in
\advance\topmargin by -0.25in
\draft

\newcommand{\bx}{{\bf x}}
\newcommand{\bz}{{\bf z}}
\newcommand{\bq}{{\bf q}}
\newcommand{\bu}{{\bf u}}
\newcommand{\bn}{{\bf n}}
\newcommand{\br}{{\bf r}}
\newcommand{\bk}{{\bf k}}
\newcommand{\bR}{{\bf R}}
\newcommand{\bQ}{{\bf Q}}
\newcommand{\bN}{{\bf 0}}
\newcommand{\bK}{{\bf K}}
\newcommand{\bG}{{\bf G}}

\newcommand{\tR}{\tilde R}
\newcommand{\tV}{\tilde V_L}

\newcommand{\erf}{\mbox{erf}}
\newcommand{\erfc}{\mbox{erfc}}
\newcommand{\ds}{\displaystyle}
\newcommand{\cH}{{\cal H}}
\newcommand{\bphi}{\bbox{\phi}}

\title{Nonuniversal Quasi-Long Range Order in the Glassy Phase of
  Impure Superconductors}

\author{Thorsten Emig, Simon Bogner and Thomas Nattermann}
\address{Institut f\"ur Theoretische Physik, Universit\"at zu K\"oln,
  Z\"ulpicher Str. 77, D-50937 K\"oln, Germany}

\date{\today}

\maketitle

\begin{abstract}
  The structural correlation functions of a weakly disordered
  Abrikosov lattice are calculated for the first time in a systematic
  RG-expansion. In the asymptotic limit the Abrikosov lattice exhibits
  still quasi long range translational order described by a {\it
    nonuniversal} exponent $\overline {\eta }_{\bf G}$ which depends
  on the ratio of the renormalized elastic constants $\kappa
  =\tilde{c}_{66}/\tilde {c}_{11}$ of the flux line (FL) lattice.  Our
  calculations show clearly three distinct scaling regimes
  corresponding to the Larkin, the manifold and the asymptotic Bragg
  glass regime. The manifold roughness exponent $\zeta_{\rm
    rm}(\kappa) $ is also {\it nonuniversal}.  Our results are
  at variance with those of the variational treatment with replica
  symmetry breaking which allows in principle an experimental
  discrimination between the two approaches.
\end{abstract}


\begin{multicols}{2}

Conventional type-II superconductors show in addition to the flux
repulsing Meissner state a second superconducting (Abrikosov) phase in
which the magnetic induction ${\bf B}$ enters the material in the form
of quantized flux lines (FLs) which form a triangular lattice.  Each
FL carries a unit flux quantum $\phi_0= hc/2e$. The Abrikosov lattice
is characterized by a non--zero shear modulus $c_{66}$, which vanishes
at the upper and lower critical fields, $H_{c_2}$ and $H_{c_1}$, where
continuous transitions to the normal and the Meissner state,
respectively, occur.  Abrikosov in his mean--field solution treats FLs
as stiff rods.
  
Thermal fluctuations roughen the FLs resulting in a melting of the
Abrikosov lattice close to $H_{c_1}$ and $H_{c_2}$, respectively,
because of the softening of $c_{66}$.  This applies in particular to
high--$T_c$ materials with their elevated transition temperatures and
their pronounced layer structures \cite{Blatter+94}.  At present, it
is not clear, whether the transition to the normal phase at high field
happens in these materials via one or two transitions.  However,
melting of the FL lattice has clearly been observed experimentally
\cite{Cubitt+93,Zeldov+95}.
  
It is well known, that in addition to thermal fluctuations in type--II
superconductors the effect of disorder has to be taken into account
since FLs have to be pinned in order to prevent dissipation from their
motion under the influence of an external current.  Randomly
distributed pinning centers lead indeed to a destruction of the
Abrikosov lattice, as has been first demonstrated by Larkin using
perturbation theory\cite{Larkin70}.  In particular, he found an {\it
  exponential} decay of the correlations of the order parameter for
translational long range order $\Psi_{\bf G}(\bf x) = e^{i{\bf
    Gu(x)}}$ on length scales larger than a disorder dependent Larkin
length $L_{\xi }$. Here ${\bf G}$ and ${\bf u}$ denote a reciprocal 
lattice vectors and the displacement field of the
FL lattice, respectively, and  $\bf x=({\bf x}_{\perp },z)$ is a 
d-dimensional position vector.
  
However, as was first shown by Nattermann \cite{Nattermann90}, in
treating the interaction between the FL lattice and the disorder, it
is crucial to keep the periodicity of this interaction under the
transformation ${\bf u} \rightarrow {\bf u}+{\bf R}$, where ${\bf R}$
is a vector of the Abrikosov lattice. This symmetry, which is
abandoned in perturbation theory \cite{Larkin70} and in the so-called
manifold models \cite{Bouchaud+92c}, leads to a much slower,
logarithmic increase of the elastic distortions with the system size
$L$ \cite{Nattermann90,Korshunov93,Giamarchi+94_95}. This results in a
power law decay of the pair correlation function $C_{\bf G} ({\bf
  x})=\langle\Psi_{\bf G}({\bf x})\Psi_{-\bf G}({\bf 0})\rangle$.  In
particular, Giamarchi and Le Doussal \cite{Giamarchi+94_95} calculated
$C_{\bf G} ({\bf x})$ using (i) a variational treatment for the
triangular FL lattice and (ii) a functional renormalization group
(FRG) for a simplified model using a scalar displacement field $u$
only.  In both cases they found $C_{\bf G} ({\bf x}_{\perp},0)\sim
|{\bf x_{\perp}}|^{-\bar \eta_{\bf G} }$ with $\bar{\eta}_{\bf G_0}
=A(4-d)$ where $d$ denotes the space dimensionality. ${\bf G}_0$ is
one of the smallest reciprocal lattice with $G_0a=4\pi/\sqrt{3}$,
$a=(2\phi_0/\sqrt{3}B)^{1/2}$ denotes the lattice spacing and $A$ is a
universal constant ($A=1$ and $A=\pi^2/9\approx 1.1$ for the treatment
(i) and (ii), respectively).  Thus, even with (weak) disorder there is
a quasi-long range ordered flux phase which shows Bragg peaks, the
"Bragg-glass" \cite{Giamarchi+94_95}.  This result is believed to be
valid in an extended region of the $H-T$ phase diagram up to a phase
boundary, where the occurrence of unbounded dislocation loops -
ignored so far - leads to an instability of the Bragg glass phase
\cite{Kierfeld+97,Gingras+96,Carp+96,Ertas+96,Kierfeld98,FisherDS97}.

The power law decay of $C_{\bf G} ({\bf x})$ is reminiscent of the situation
in pure $2D$-crystals where in the solid phase $C_{\bf G} ({\bf
  x_{\perp}})\sim |{\bf x_{\perp}}|^{-\eta_{{\bf G}}}$.  This solid
phase is indeed a {\it line} of critical points with $\eta_{{\bf
    G}}=TG^2(1+{\kappa}^{-1})/({4\pi}{\tilde c}_{11})$ and $\kappa =
{\tilde c}_{66}/{\tilde c}_{11}$. The ${\tilde c}_{ii}$ represent the
{\it renormalized} elastic constants which have a finite temperature
dependent value. {\it At} the melting temperature $T_m$ 
%$\eta_{\bf{G}}$ reaches a nonuniversal value 
$\eta_{\bf{G}_0}={1 \over 3}(1-{\kappa ^2})$.
%Which still depends on $\kappa(T_m)$
\cite{Nelson83}. Similarly, the $2D$ disordered crystal close to the glass temperature $T_g$ (i.e. $\eta_{\bf{G}_0} \lesssim 2$) is described by a line of fixed points which depends on $\kappa$ \cite{Carp+97}.
  
On the contrary, the Bragg glass phase is characterized by a zero
temperature fixed point where the ratio between the disorder strength
and the elastic energy vanishes on large scales $L$ as
$\Delta^*L^{d-4}$. It has been suggested in \cite{Giamarchi+94_95},
that the fixed point value $\Delta^*$ might be {\it universal}
resulting in a universal coefficient $A$ of $\overline {\eta}_{\bf
  G}$.  In the present treatment we will show however, that a FRG
treatment of the {\it full } triangular model leads indeed to a
nonuniversal {\it $\kappa$-dependent} value of $\overline{\eta}_{\bf
  G}=\Delta^*(\kappa)(Ga)^2$.  The situation is therefore
qualitatively similar to that of $2D$ crystals {\it at} the melting
temperature.  Quantitatively, the effect reflects the contributions
from the interaction between different Fourier modes, which are not
considered in the variational treatment
\cite{Korshunov93,Giamarchi+94_95}.  With $\kappa$ dependent in
general on ${\bf B}$ and $T$, the observation of a field-dependent
$\bar\eta_{\bf G}$ would yield the opportunity to judge the validity
of different approximation schemes under debate \cite{Balents+96}.

Since in the Bragg-glass phase dislocations in the vortex lattice can
be neglected
\cite{Kierfeld+97,Gingras+96,Carp+96,Ertas+96,Kierfeld98,FisherDS97},
its configurations are described completely by the elastic
displacement field $\bu(\bx)$. The vortex density is then given by
$\rho(\bx,\bu)=\sum_\bR \delta (\bx_\perp-\bR-\bu(\bR,z))$.  The
impurities are modeled by a Gaussian random potential $U(\bx)$ with
two-point correlation $\overline{U(\bx)U(\bN)}= \gamma\xi^2
g(x_\perp/\xi)\delta(z)$. Here $\gamma =f_{\rm pin}^2 n_i \xi^2 $,
$f_{\rm pin}$ denotes the individual pinning force, $n_i$ the impurity
density, $\xi$ the maximum of the coherence and disorder correlation
length \cite{Blatter+94} and $g$ is a short ranged function.
On scales larger than the London penetration depth $\lambda$ the
FL lattice can be described by local elasticity theory, leading to the
Hamiltonian
\begin{eqnarray}
\label{Ham}
\cH&=&\frac{1}{2}\int d^{d-2}\bz d^2 \bx_\perp\left\{
c_{11}\left(\bbox{\nabla}_\perp\cdot\bu\right)^2+
c_{66}\left(\bbox{\nabla}_\perp\times\bu\right)^2+
\right.\nonumber \\
&+& \left. c_{44}\left(\bbox{\nabla}_\bz
  \bu\right)^2  \right\} + \int d^d\bx E_{\rm pin}(\bu,\bx),
\end{eqnarray}
where $\bx_\perp=(x,y)$, and the $\bz$ component has been generalized
to a $d-2$-dimensional coordinate to allow for an $\epsilon$-expansion
about 4 dimensions.  The random potential is defined as $E_{\rm
  pin}(\bu,\bx)=U(\bx)\rho(\bx,\bu)$.  The Hamiltonian is invariant
under simultaneous mapping of shear onto compressional modes (rotation
of $\bu$ by $90^{\circ}$) and permutation of the corresponding moduli
($\kappa \rightarrow \kappa^{-1}$). For $E_{\rm pin}(\bu,\bx)$ this
invariance holds in a statistical sense. Thus the partition function
and all thermodynamic quantities also show this symmetry. We therefore
can restrict the calculations to $0 \le \kappa \le 1$ in the
following.  Results for $\kappa > 1$ correspond to those for
$\kappa^{-1}$ with $\bu$ rotated.

Performing the disorder average and discarding spatially rapidly
oscillating terms \cite{Giamarchi+94_95}, one obtains for the disorder
correlator, which is defined by $\overline{E_{\rm pin}(\bu,\bx)E_{\rm
    pin}(\bN,\bN)}=R(\bu)\delta(\bx)$, a periodic function with the
Fourier representation
\begin{equation}
    \label{R_bare}
    R(\bu)=
\gamma\xi^2\frac{B^2}{\phi_0^2}
    \sum_\bG \hat g(G\xi) \cos(\bG\bu),
\end{equation}
where $B/\phi_0$ is the mean area density of vortices and
$\hat g(G\xi)$ is the Fourier transform of $g(x_\perp/\xi)$. 

To obtain the disorder averaged configuration of the FL lattice on a
particular length scale, one has to take into account the
renormalization of $R(\bu)$ by fluctuations on shorter length scales.
This can be done systematically by a functional renormalization group
(FRG) for the replica Hamiltonian resulting from Eq. (\ref{Ham}) after
the disorder average. Because of the statistical invariance of the the
replica Hamiltonian with the inter-replica coupling (\ref{R_bare})
under a shift of $\bu$ by an arbitrary vector field inducing a
compression, shear and/or tilt of the FL lattice, there is no renormalization
of the elastic moduli \cite{Hwa+94}.  Therefore the temperature obeys
the exact flow equation $dT/dl=(d-2)T$ leading to a $T=0$ fixed point
for $d>2$.  Notice however that in the original model of (\ref{Ham})
the statistical invariance is not exactly fulfilled on length scales
smaller than $a$, leading to a small renormalization $c_{ii}
\rightarrow \tilde{c}_{ii}$ of the elastic constants which will be
considered from now on as effective parameters.  In our FRG only
coordinates are rescaled as $\bx \to \exp(dl)\bx$ to keep the cutoff
$\Lambda$ fixed with $dl$ the infinitesimal width of the momentum
shell.  Because of the dispersion of the elastic constants on scales
smaller than the penetration depth $\lambda$, we have to choose here
$\Lambda\approx 2\pi/\lambda$. Fluctuations on smaller scales can be
ignored if the Larkin length $L_{\xi}$ (see below) is much larger then
$\lambda$, i.e. for weak disorder. For larger disorder one has to take
into account the dispersion $\tilde{c}_{11}$ and $\tilde{c}_{44}$
which will result in a more complicated cross-over but not affect the
asymptotic behavior.  The flow equation for $R(\bu)$ can then be
derived along the lines of Refs. \cite{FisherDS86a,Balents+93}.
Contrary to previous cases, after suitable rescaling of $R(\bu)$, we
obtain
\begin{eqnarray}
  \label{RGflow}
&&\frac{d R(\bu)}{dl} = \epsilon R(\bu) +\frac{a^2}{2}\bigg\{\!
\left(\partial_x^2 R\right)^2+\left(\partial_y^2 R\right)^2+2\left(
  \partial_x\partial_y R\right)^2\nonumber\\
&+&2\Delta\left(\partial_x^2 + \partial_y^2\right)\!\! R
-\frac{\delta}{4}\left[\!\left(\partial_x^2 R-\partial_y^2 R\right)^2\!+
  4\left(\partial_x \partial_y R\right)^2\right]\!\!\bigg\}\!
\end{eqnarray}
with the dimensionless parameter $\Delta=-\partial_x\partial_x
R(\bN)\equiv -\partial_y\partial_y R(\bN)$ and $\partial_x =
\partial/\partial u_x$ etc.. The last equality as well as
$\partial_x\partial_y R(\bN)=0$ follow from the requirement of
hexagonal symmetry for $R(\bu)$. Performing the momentum shell
integrals to lowest order in $\epsilon$, one obtains the anisotropy
parameter
\begin{equation}
\delta=1-\frac{2\ln(\kappa)}{\kappa-\kappa^{-1}},
\end{equation}
i.e., $0\le \delta \le 1$ for any ratio
$\kappa=\tilde{c}_{66}/\tilde{c}_{11}$.  In the isotropic case $\tilde
c_{11}=\tilde c_{66}$ ($\delta=0$) the flow Eq. (\ref{RGflow}) reduces
to that of Ref. \cite{Balents+93}, if $\tilde c_{11}\rightarrow
\infty$, as often assumed for FL lattices, we have to put $\delta=1$.

{From} Eq. (\ref{R_bare}) we find that the bare, unrenormalized
dimensionless parameter $\Delta_0$ is given by
\begin{equation}
  \label{bare_Delta}
  \Delta_0
\approx 10^{-3}\Lambda^{d-4} \gamma\frac{1+\kappa}{\tilde c_{44}\tilde c_{66}}
\frac{B^2}{\phi_0^2}\left(1+{\cal O}(\xi/a)\right), 
\end{equation}
where we have assumed $\hat g(G\xi)=\xi^2\Theta(1-G\xi)$ with $\Theta$
the step function and $\xi \ll a$ to evaluate the sum over $\bG$.  The
Larkin length (parallel to ${\bf B}$) in three dimensions is given by
$L_{\xi}\approx \xi^2/\Delta_0\Lambda a^2$ \cite{Blatter+94}. The condition
of weak disorder mentioned above reads then $\Delta_0a^2\ll \xi^2$. \nopagebreak
 
Now we integrate Eq. (\ref{RGflow}) to obtain the renormalized
function $R(\bu)$ on all length scales $L=\exp(l)/\Lambda$ including
the fixed point $R^*(\bu)$ for $L\to \infty$. With the bare corre- \pagebreak
\begin{figure}[tb]
\begin{center}
\leavevmode
\epsfxsize=0.75\linewidth
\epsfbox{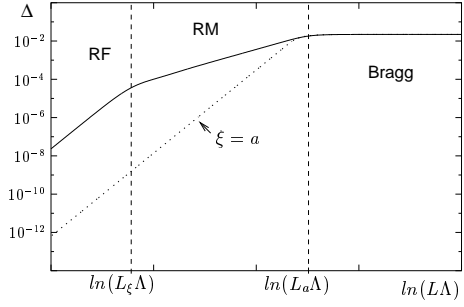}
\end{center}
\caption{RG flow of $\Delta(l)$ through three regimes obtained by
numerical integration of Eq. (\ref{RGflow}).}
\label{fig_co}
\end{figure}
\noindent lator of Eq. (\ref{R_bare}) showing the full symmetry of the
triangular lattice -- translation, 6-fold rotational axis, 3 mirror
lines -- and the flow of Eq. (\ref{RGflow}) preserving these
symmetries as it ought to, the set of possible solutions is restricted
to functions with the full lattice symmetry on every length scale.
Even for $R^*(\bu)$ an analytic solution not being very obvious, our
approach is numerical and straightforward: The Ansatz assuring
translational plus inversion invariance is $R(\bu)=\sum_\bG R_\bG
\cos(\bG \bu)$ with the sum running over the reciprocal lattice.  An
infinite set of coupled nonlinear but ordinary flow equations for the
coefficients $R_\bG$ is derived.  Rather than solving directly for the
fixed point $R^*_\bG$, the flow equations are integrated numerically
with the remaining point group symmetries exploited here. Convergence
to a fixed point from a huge bassin of attraction is observed. Of
special interest is the flow of $\Delta(l)$ since it determines the
scale dependence of the renormalized propagator $\Delta_{\rm
  eff}(q)/q^4$ with $\Delta_{\rm
  eff}(q)=\Delta(\ln(\Lambda/q))(q/\Lambda)^{4-d}$ and thus the
exponents $\bar\eta_\bG$ and $\zeta_{rm}$.  Obviously, the numerical
integration has to be restricted to a {\it finite} set of $R_\bG$ with
$|\bG|<|\bG_c|$, but high accuracy of the flow of $\Delta(l)$ can be
restored by including up to 360 different coefficients and a finite
size scaling like extrapolation.  The accuracy for $L\to\infty$ is
controlled via the exact ($\delta$-independent) relation $\epsilon
R(\bN)=(a\Delta)^2$ stemming from (\ref{RGflow}).

\begin{figure}[tb]
\begin{center}
\leavevmode
\epsfxsize=0.8\linewidth
\epsfbox{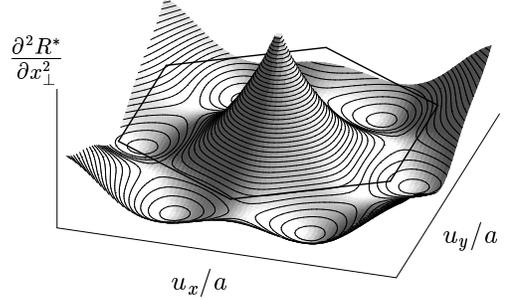}
\end{center}
\caption{Characteristic of the Bragg glass: a cusp in 
$\partial_{x_\perp}^2 R^*(\bu)$ at the lattice sites.}
\label{fig_fp}
\end{figure}

\begin{figure}[tb]
\begin{center}
  \leavevmode \epsfxsize=1.0\linewidth \epsfbox{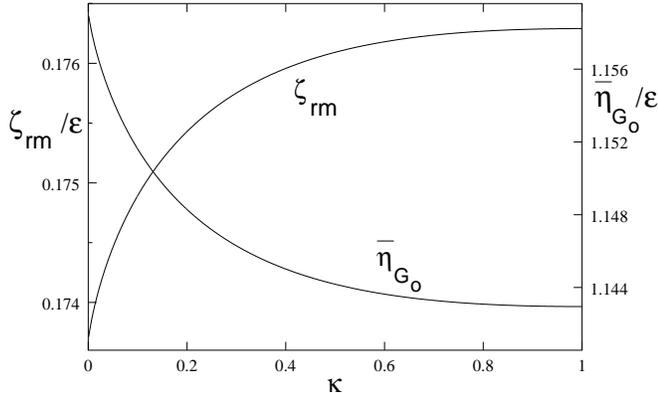}
\end{center}
\caption{Variation of the exponents $\zeta_{\rm rm}$ and $\bar\eta_{\bG_0}$ 
         with $\kappa$.}
\label{fig_exp}
\end{figure}

Two different but typical kinds of behavior of $\Delta(l)$ are shown
in Fig. \ref{fig_co}.  Depending on the ratio $\xi/a$, two or three
scaling regimes with different roughness exponent $\zeta$ defined by
$\Delta(l)\sim \exp(2\zeta l)$ can be clearly identified. For
$\xi/a\ll 1$, one finds a first crossover from the random force (RF)
regime ($\zeta=\epsilon/2$) to the random manifold (RM) regime
($\zeta=\zeta_{\rm rm}$) at the Larkin length $L_{\xi}$ where $u
\simeq \xi$ and a second one at the positional correlation length
$L_a\approx L_{\xi}(a/\xi)^{1/\zeta_{rm}}$ where $u \simeq a$ to the
Bragg glass regime ($\zeta=0$). If $\xi/a\simeq 1$, $L_a\to L_{\xi}$
and the RM regime disappears.  Just like in the 1-d (or CDW) case
\cite{Giamarchi+94_95} a cusp in the second derivatives of $R(\bu)$
develops leading to the fixed point shape shown in Fig. \ref{fig_fp}.
The length scale $L_{\rm c}$ on which the cusp appears can be
determined analytically from the divergence in the RG flow of
$\partial_x^4 R(\bN)=\partial_y^4 R(\bN)=3\partial_x^2\partial_y^2
R(\bN)$. As can be expected, $L_{\rm c}$ is related to the Larkin
length $L_\xi$ known from perturbation theory by a numerical factor
only. But interestingly we obtain the typical ratio $L_{\rm
  c}/L_\xi\approx 100$, i.e., the cusp appears clearly beyond the
Larkin scale.  The {\it nonuniversality} of the exponents $\zeta_{\rm
  rm}(\kappa)$ and $\bar\eta_{\bG}(\kappa)$ can be determined from
$\Delta(l)$ via the slope in the RM regime and $\Delta^*$,
respectively.  Their dependence on $\kappa$ is presented in Fig.
\ref{fig_exp}.
With the numerical value for $\Delta^*(\kappa)$ at hands we can now
calculate the displacement correlations
$B_{ab}(\bx)=\overline{\langle[u_a(\bx)-u_a(\bN)][u_b(\bx)-u_b(\bN)]\rangle}$
in the Bragg glass phase. Introducing the rescaled $z$-coordinates
$z_t=(\tilde c_{66}/\tilde c_{44})^{1/2}z$, $z_l=(\tilde
c_{11}/\tilde c_{44})^{1/2}z$ and $h(t)=t^{-2} \, ln(1+t^2)$, we find for $x \gg L_a$ \beginwide
\begin{eqnarray}
\label{displace_xx}
B_{xx}(\bx)&=&\frac{\Delta^*(\kappa)a^2}{1+\kappa}\left\{
\ln\!\left(\frac{x_\perp^2+z_{t}^2}{L_a^2}\right)
+\kappa \ln\!\left(\frac{x_\perp^2+z_{l}^2}{L_a^2}\right)
+\frac{y^2-x^2}{x_\perp^2}\left[1-\kappa-h\!\left(\frac{x_\perp}{z_t}\!\right)+\kappa \, h\!\left(\frac{x_\perp}{z_l}\!\right)\right]\right\}\\
\label{displace_xy}
B_{xy}(\bx)&=&\frac{2\Delta^*(\kappa)a^2}{1+\kappa}\frac{xy}{x_\perp^2}
\left\{\kappa-1-\kappa \, h\!\left(\frac{x_\perp}
{z_l} \right)+h\!\left(\frac{x_\perp}
{z_t} \right)\right\},
\end{eqnarray}
\endwide
and $B_{yy}(\bx)$ follows from $B_{xx}(\bx)$ by permuting
$\tilde c_{11}$ and $\tilde c_{66}$. These correlations lead to the
translational order correlation function $C_\bG(\bx)=\overline{\langle
\exp(i\bG[\bu(\bx)-\bu(\bN)]) \rangle}$, which reads
\begin{equation}
\label{CK-corr}
C_\bG(\bx)  \sim  g_\bG L_{a}^{\bar\eta_\bG}(x_\perp^2+z_{t}^2)^
{-\bar\eta_\bG/(2(1+\kappa))}
(x_\perp^2+z_{l}^2)^{-\bar\eta_\bG/(2(1+1/\kappa))}
\end{equation}
with $\bar\eta_\bG=\Delta^*(\kappa)(aG)^2$ and the geometrical factor
\begin{eqnarray}
 && g_\bG  =  \exp\left[\frac{\Delta^*(\kappa)(aG)^2}{1+\kappa}\left(
      (\hat\bx_\perp\hat\bG)^2-\frac{1}{2}\right)\! \, \right. \\ 
 &&\times \, \left.
    \left\{\!\left(\!1- h\!\left(\frac{x_\perp}{z_t}\right)\!\!\right)
      -\kappa\left(\!1-h\!\left(\frac{x_\perp}{z_l}\right)\!\!\right)\!\right\}
         \right], \nonumber
\end{eqnarray}
which describes completely the angular dependencies of the
translational order. Note that the factor $g_\bG$ goes to $1$ in the
limit $z\to\infty$. Therefore, in this limit the dependence of
$C_\bG(\bx)$ on the reciprocal lattice vector $\bG$ remains only in
the exponent $\bar\eta_\bG$. Moreover, it is interesting to note that
the exponents of the algebraic decay in Eq. (\ref{CK-corr}) depend on
the elastic moduli as soon as $z$ is finite even without taking into
account the nonuniversality of the exponent $\bar\eta_\bG$ itself.
Neglecting the non-trivial renormalization of $\Delta^*(\kappa)$, in
the case $z=0$ the above formulae reduce to those found in
\cite{Giamarchi+94_95}.

The $\kappa$-dependence of $\bar\eta_{\bG_0}$ and $\zeta_{rm}$ (Fig
3), the correlation functions $B_{ab}(\bx)$ (Eqs.(6),(7)) and
$C_{\bG}(\bx)$ (Eqs.(8),(9)) and the cross-over from the Larkin via the
random manifold to the Bragg glass regime (Fig. 1) are the main results
of this paper.

In isotropic superconductors at low temperatures, where flux lines
interact via central forces, on has $0\le\kappa\le 1/3$. $\kappa\sim
1/3$ for $\lambda\le a$, i.e. for fields close to $H_{c_1}$, and
$\kappa \rightarrow 0$ for $H \rightarrow H_{c_2}$.  For most of the
field region $\kappa \approx \phi_0/16\pi\lambda^2B$,.  Thus, an
increase of the external field from $H_{c_1}$ to $H_{c_2}$ should
result in an increase of $\bar\eta_{{\bf G}}$ and a decrease of
$\zeta_{rm}$. Numerically, the effect is small, since $\bar\eta_{{\bf
    G}_0}$ ranges from $1.143$ to $1.159$ and $\zeta_{rm}$ from
$0.1737$ to $0.1763$ in this $\kappa$ range. Thus it will be probably
hard to detect this effect.  At higher temperatures, where the flux
line interaction is considerable influenced by thermal interaction, as
well as in anisotropic superconductors, the above inequality for
$\kappa$ may not be longer fulfilled. Clearly, in the latter case also
our starting Hamiltonian (1) would have to be modified.

Contact to the neutron scattering experiment is made by 
%\begin{equation}
$S(\!\bG\!+\! \bq) \! = \!   \int \!d^3x e^{i \bq \bx} C_\bG(\bx)$,
%\end{equation}
the structure factor, which is proportional to the scattered intensity.
% It describes the divergence of the scattered intensity with vanishing $\bq$.
For the marginal cases
$\kappa=0$ and $\kappa=1$ we find from Eq.  (\ref{CK-corr})
\begin{equation}
S(\bG+\bq)\sim \left(\bq_\perp^2+\frac{\tilde c_{44}}{\tilde c_{66}}
q_z^2\right)^{(-3+\bar\eta_\bG(\kappa))/2},
\end{equation}
i.e., $S(\bG+\bq)$ exhibits Bragg peaks. For $\kappa>0$ the Fourier transform  can be easily done numerically.
% from Eq. (\ref{CK-corr}).

To summarize, we have shown that contrary to previous claims, the
translational quasi-long range order in the Bragg glass phase of
impure type-II superconductors is described by a nonuniversal
power-like decay of the order parameter correlations. In particular,
the decay-exponent $\bar\eta_{\bf G}$ depends on the ratio
$\kappa={\tilde c}_{66}/{\tilde c}_{11}$ of the elastic constants,
similar to $2D$ pure crystals at their melting temperature. For weak
disorder we find a crossover of the structural correlation functions
from a Larkin-regime, where perturbation theory applies, to the random
manifold regime and eventually to the asymptotic Bragg glass regime.
This nonuniversality could be in principle tested by neutron
scattering changing the external field.

We acknowledge discussions with H.E. Brandt, D. Feldman and S. Scheidl. 
This work was supported by DFG through SFB 341.

\end{multicols}{}
\end{document}